\documentclass[twocolumn,amsmath,amssymb,aps,prb]{revtex4-1}

\usepackage{times}
\usepackage{amsmath}
\usepackage{amssymb}
\usepackage[]{graphicx}
\usepackage{bm}
\usepackage{color}
\usepackage{soul}
\usepackage{float}
\usepackage{hyperref}
\begin{document}

\title{Rotation Rate of Rods in Turbulent Fluid Flow}

\author{Shima Parsa$^{1}$, Enrico Calzavarini$^{2}$, Federico Toschi$^{3}$, and Greg A. Voth$^{1}$$^\ast$\\
$^{1}$Department of Physics, Wesleyan University,\\ Middletown, Connecticut 06459, USA\\
$^{2}$Laboratoire de M\'ecanique de Lille CNRS/UMR 8107,\\ Universit\'e Lille 1, 59650 Villeneuve d'Ascq, France\\
$^{3}$Department of Physics, and Dept. of Mathematics \& Computer Science,\\
 Eindhoven University of Technology, 5600 MB Eindhoven, The Netherlands\\
 $^\ast$ Correspondence to: gvoth@wesleyan.edu
}

\begin{abstract}
The rotational dynamics of anisotropic particles advected in a turbulent fluid flow are important in many industrial and natural setting.  Particle rotations are controlled by small scale properties of turbulence that are nearly universal, and so provide a rich system where experiments can be directly compared with theory and simulations.  Here we report the first three-dimensional experimental measurements of the orientation dynamics of rod-like particles as they are advected in a turbulent fluid flow.  We also present numerical simulations that show good agreement with the experiments and allow extension to a wide range of particle shapes.  Anisotropic tracer particles preferentially sample the flow since their orientations become correlated with the velocity gradient tensor.  The rotation rate is heavily influenced by this preferential alignment, and the alignment depends strongly on particle shape.

\end{abstract}

\date{\today}

\maketitle
The dynamics of anisotropic particles in turbulent fluid flows are central to understanding many applications ranging from cellulose fibers used in paper making~\cite{Lundell2011} to ice crystals in clouds~\cite{sherwood2006,pinsky1998} to locomotion of many micro-organisms~\cite{Pedley1992,Bowen1993,Saintillan2007,Locsei2009}.    When the particles are small and their concentration is low, their rotations are determined by the velocity gradient along their trajectories.  Velocity gradients are dominated by the smallest scales in turbulence about which we have an extensive fundamental understanding~\cite{Zeff2005,Luthi2005,Meneveau2011}.  So the statistics of rotations of anisotropic particles forms an important problem foundational to many applications for which we need to develop a predictive understanding based on the fundamental properties of turbulence.

Advances in imaging technology have made it possible to obtain time-resolved trajectories of particles in turbulence using high speed stereoscopic imaging, and these measurements have produced many new insights about Lagrangian translational dynamics of spherical particles~\cite{Toschi2009}.   There is also an extensive literature on the motion of anisotropic particles in fluid flows.  Anisotropic particles have fascinating dynamics even in simple flows~\cite{jeffery1922,Szeri1992,Wilkinson2009,Gavze2011}.   Several simulations have addressed the turbulent case~\cite{Shin2005,Mortensen2008,Pumir2011,Zhang2001,Marchioli2010}, and experiments have measured dynamics in 2D flows~\cite{Parsa2011} and orientation distributions in 3D turbulent flows~\cite{Bernstein1994,Newsom1998,Parsheh2005}.  However, there are no available experimental measurements of time-resolved rotational dynamics of anisotropic particles in 3D turbulence.

When ellipsoidal particles are small compared with the smallest length scales in the flow, their rotation rate is determined by the velocity gradient tensor through Jeffery's equation~\cite{jeffery1922}:
\begin{equation}
\dot{p}_i = \Omega_{ij} p_j+\frac{\alpha^2-1}{\alpha^2+1}(S_{ij} p_j-p_i p_k S_{kl}p_l)
\label{Eq:Jeffrey}
\end{equation}
where $p_i$ is a component of the orientation director and $\alpha\equiv l/d$, is the aspect ratio of the ellipsoid given by the ratio of length ($l$) to diameter ($d$).  $\Omega_{ij}$ is the rate-of-rotation tensor, and $S_{ij}$ is the rate-of-strain tensor which are the anti-symmetric and symmetric parts of the velocity gradient tensor respectively.  Particle dynamics become much more complicated when the particles are large~\cite{Shin2005} or density mismatched~\cite{sherwood2006}, or high particle concentration produces inter-particle interactions and two-way coupling~\cite{Lundell2011}.  A phenomenology of the dynamics of anisotropic tracers at low concentrations in turbulence is needed to provide a foundation for continued work on the more complex cases that often appear in applications.

We have performed an experimental and computational study of rotation rates along the trajectories of anisotropic particles in turbulent flows. In the experiments, small rods are tracked using stereoscopic images from four high speed cameras.
Turbulence was generated between two grids oscillating in phase in an octagonal Plexiglas tank \cite{Blum2010}
that is $1\times1 \times $1.5 m$^3$. The experiments are performed at two Taylor Reynolds number, $R_\lambda = (15uL/\nu)^{1/2}$ , $R_\lambda=$160 and  $R_\lambda=$214.
The Kolmogorov length scale ($\eta = (\nu^3/\langle\epsilon\rangle)^{1/4}$) and time scale ($\tau_\eta =(\nu/\langle\epsilon\rangle)^{1/2} $) are respectively $\eta =375 \mu $m and $\tau_\eta =70 m$s,
and $\eta =210 \mu $m and $\tau_\eta =25 m$s. The rods are 1~mm in length by 200$\mu $m in diameter and are cut from nylon thread at density of  $\rho=1.15 $ g/cm$^3$. The fluid is density matched with rods by adding 19\%  CaCl$_2$ by mass to water. The position of the center of the rods was measured with an uncertainty of  $\approx 40 \mu $m. We have measured the orientation of rods in two dimensional (2D) images from each camera and reconstructed the orientation of rods
in three dimensional (3D) space using images from multiple cameras.
Determining the orientation of the rods in 3D requires measurements of the orientation in 2D images from at least three cameras that are not in the same plane.The rotation rate, $\dot{\mathbf{p}}$, is determined from quadratic fits to the measured orientation versus time data. We have also studied the motion of anisotropic tracers in direct numerical simulations (DNS) of homogeneous and isotropic turbulence at  $R_\lambda =$180.   The translational motion of tracer rods matches that of fluid particles, so we are able to use a database of previously simulated Lagrangian trajectories~\cite{Benzi2009} to integrate Jeffery's equation~(\ref{Eq:Jeffrey}) and obtain the time evolution of particle orientations. The spatial spectral resolution was $512^3$ points. These simulations stored the full velocity gradient tensor along Lagrangian trajectories at time intervals of about $1/10 \tau_{\eta}$.  Equation~\ref{Eq:Jeffrey} has been therefore integrated \textit{a posteriori} with the same integration time step and with an Adams-Bashforth second order in time scheme.
\begin{figure}[tb]
 \centerline{\includegraphics[scale=1]{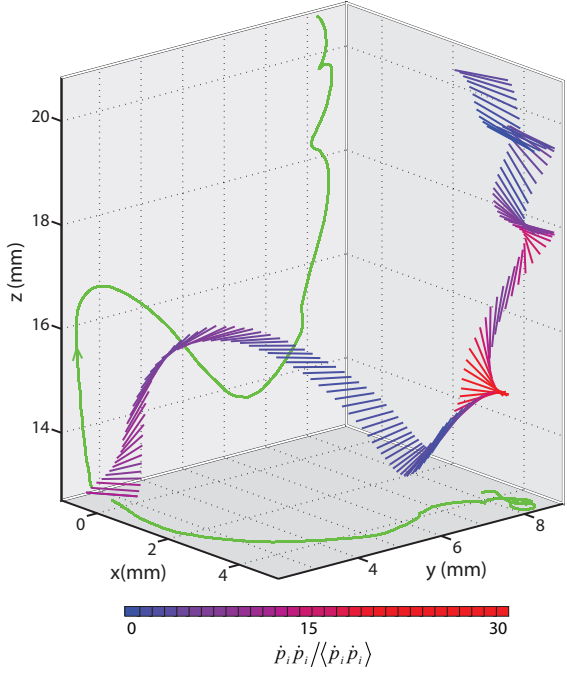}}
\caption{ Three-dimensional view of a rod trajectory with a large rotation rate from the experiment at $R_\lambda=214$. The color of the rod represents the rotation rate.  This rod is tracked for  284~ms.  The green lines show the projection of the center of the rod onto the y-z and x-y planes. The rod is a circular cylinder with length 1~mm and diameter 0.2~mm.
}
\label{Fig:plot3D}
\end{figure}

Figure~\ref{Fig:plot3D} shows an experimentally measured trajectory of a 1mm rod at $R_\lambda=214$.  This example illustrates several of the important properties of the rotation of rods.  First, this rod has bursts of high rotation rate where the rotation rate squared is up to 30 times its mean reflecting the intermittency of rod rotations.  Second, in the upper right, the rod is caught in a vortex, but its rotation rate is not large because the rod has become aligned with the vorticity reflecting the tendency of anisotropic particles to become aligned by the velocity gradients in the flow.

\begin{figure}[tb]
 \centerline{\includegraphics[scale=1]{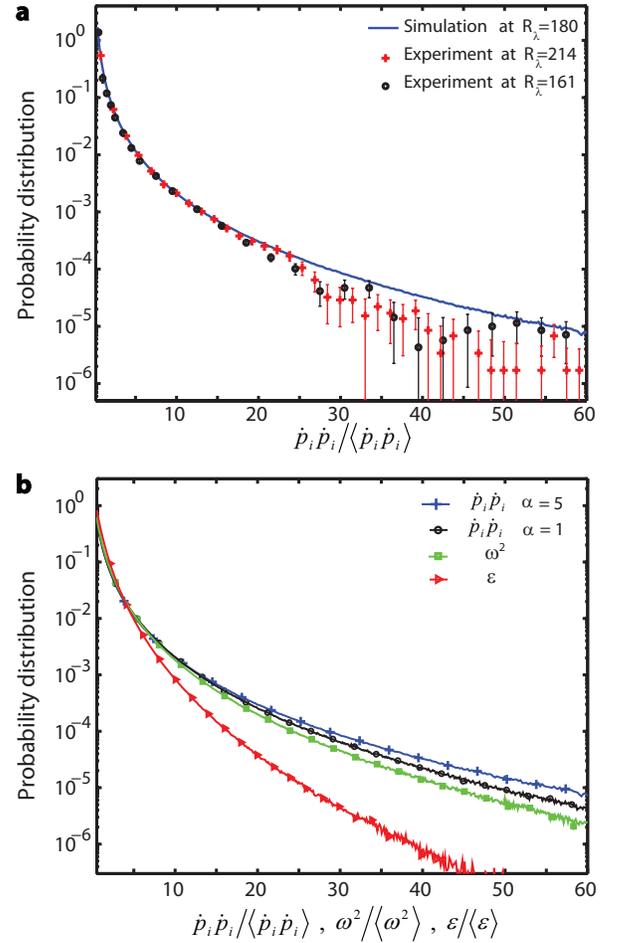}}
\caption{ The pdf of rotation rate squared of rods.
(a) Comparison of simulation (Blue line, $R_\lambda = 180$) with experiment (Black $\circ$, $R_\lambda = 160$; Red $+$, $R_\lambda = 214$) for aspect ratio $\alpha=5$. Error bars represent both random uncertainty and the systematic uncertainty produced by measuring rotation rates from experimental orientation data over a time interval.
(b) Comparison of the pdf of rotation rate squared for rods at $\alpha=5$ (Blue $+$) and spheres at $\alpha=1$ (black $\circ$) with pdfs of enstrophy (Green $\Box$) and energy dissipation rate (red $\triangleright$)
for the simulation at $R_\lambda = 180$. One symbol is displayed for every twenty bins.
}
\label{Fig:PDFp2}
\end{figure}

The probability distribution function (pdf) of the rotation rate squared, $\dot{p}_i\dot{p}_i$, of rods is shown in Fig.~\ref{Fig:PDFp2}a. The pdf has long tails which indicates the presence of rare events with
rotation rates squared up to 60 times the average value ($\langle \dot{p}_i\dot{p}_i \rangle$).
The agreement between DNS and experiment is very good for the core of the distribution up to 20~$\langle \dot{p}_i\dot{p}_i \rangle$.  At larger rotations rates the experimental pdfs are slightly below the DNS.  This difference is not much larger than the systematic errors in the experimental data represented by the error bars, but it may reflect the effect of finite length of the rods in the experiment.  The rod lengths are $2.6 \eta$ at $R_\lambda$=160 and $4.7\eta$ at $R_\lambda$=214.  Ref [16] 
indicates that rods less than about $7\eta$ should not have a measurable change in their rotation rate variance from the tracer limit, but it is possible that the rare events are more sensitive to rod length.
In Fig.~\ref{Fig:PDFp2}b, the probability distribution of rotation rate squared from the numerical simulation is compared with the distributions of enstrophy ($\omega^2=2\Omega_{ij}\Omega_{ij}$)  and
energy dissipation rate ($\epsilon=2\nu S_{ij}S_{ij}$, where $\nu$ is the kinematic viscosity).  For large values of rotation rate squared, the pdf for rods ($\alpha=5$) is larger than that of spheres ($\alpha=1$), and both are larger than either enstrophy or energy dissipation rate.  Spheres simply rotate with the vorticity, $\dot{p}_i\dot{p}_i =  \Omega_{ij}\Omega_{ij} \sin^2{\theta}$, and the larger intermittency for spheres compared with enstrophy comes from the distribution of the angle, $\theta$, between $\mathbf{p}$ and the vorticity vector.  Rods with $\alpha=5$ have a contribution to their rotation rate from $S_{ij}$, which one might think would decrease the probability of large rotations, making it more like the distribution of energy dissipation rate.  However, variations in the orientation of the rod with respect to the velocity gradient tensor contribute additional fluctuations giving rods the most high rotation rate events.

\begin{figure}[H]
 \centerline{\includegraphics[scale=1]{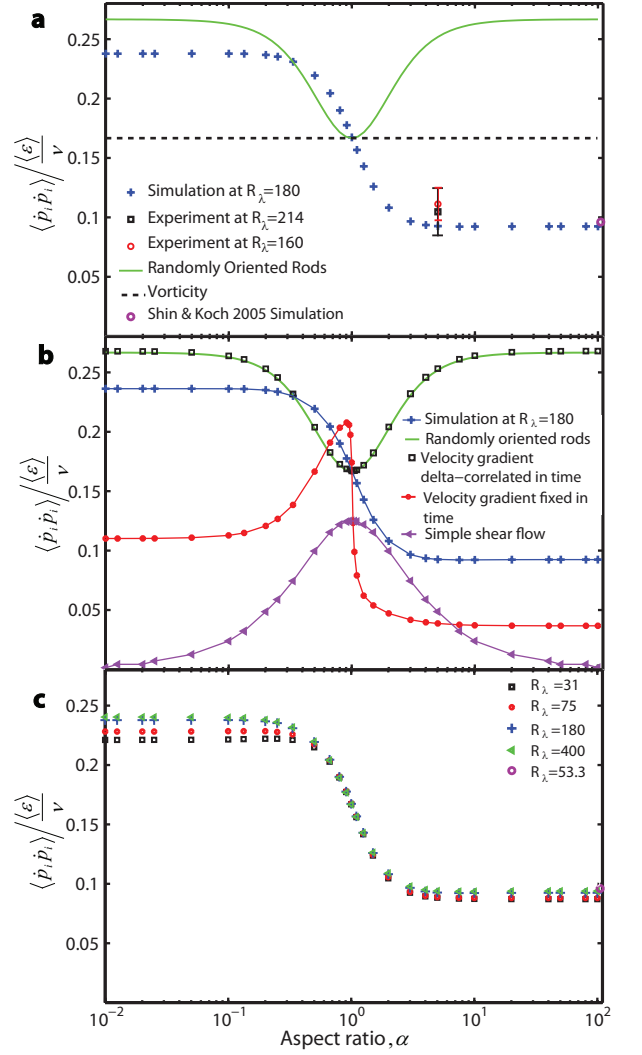}}
\caption{Rotation rate variance as a function of aspect ratio.  (a) Blue $+$ is the DNS at $R_\lambda =180$.
Black $\square$ is the experiment at $R_\lambda =214$.   Red $\circ$ is the experiment at $R_\lambda =160 $.  The error bars on the experimental points represent systematic error due to extrapolation from the finite fit time required to measure the rotation rate~\cite{Voth2002a}.  The purple $\circ$ shows the result for infinite aspect ratio rods from simulations by Shin and Koch~\cite{Shin2005} at $R_\lambda =53.3$.  The Green line is the analytic prediction for randomly oriented rods from Eq.~\ref{Eq:p2var}. The dashed line indicates the rotation due to vorticity.
(b) Simple cases of particles integrated through different velocity gradient fields:  Blue $+$ uses velocity gradients along Lagrangian trajectories at $R_\lambda =180$.   Black $\square$ uses velocity gradients sampled from the DNS and delta-correlated in time.  This data matches the analytic prediction for randomly oriented rods shown as the green line.  Red $\bullet$ uses velocity gradients sampled from the DNS and held fixed in time.  The purple $\blacktriangleleft$ uses the velocity gradient of a plane shear flow. (c) Comparison of simulations at different Reynolds numbers.
}
\label{Fig:varaincep2}
\end{figure}

Figure~\ref{Fig:varaincep2} shows the effect of the
shape of particles on their rotation rate in turbulence.  The rotation rate variance of disk shaped particles ($\alpha<1$) is much larger
than that of the spheres ($\alpha=1$).  This can be qualitatively understood as the additional contribution of strain ($S_{ij}$ in equation~\ref{Eq:Jeffrey}) to the rotation rate.  However, the rotation rate of rods ($\alpha >1$) is much smaller than spheres even though the rate-of-strain contributes to their rotation as well.  At large $\alpha$, our simulations agree with earlier work on thin rods at lower Reynolds number by Shin and Koch~\cite{Shin2005}.
The experimental measurements at $\alpha=5$ are consistent with the simulations considering the measurement uncertainties.

Understanding the rotation rate data in Fig.~\ref{Fig:varaincep2} requires considering the preferential alignment that occurs between particles and the velocity gradient tensor.  When particles are oriented randomly, their rotation rate variance can be calculated analytically from Eq.~\ref{Eq:Jeffrey} by extending the calculation in Ref [16] 
 to finite aspect ratio:
 \begin{equation}
\frac{\langle\dot{p}_i\dot{p}_i\rangle}{ \langle\varepsilon \rangle/ \nu}=\frac{1}{6}+\frac{1}{10}\left(\frac{\alpha^2-1}{\alpha^2+1}\right)^2\
\label{Eq:p2var}
\end{equation}
This result is shown as the green solid curve in Fig~\ref{Fig:varaincep2}a.  As particles are advected by the flow, they become oriented so that their rotation rates are very different than the randomly oriented case, with the largest difference occurring for thin rods ($\alpha >> 1$).  Thin tracer rods are material line segments, so the large decrease in rotation rate for thin rods can be qualitatively understood as resulting from the known preferentially alignment of material lines with the vorticity vector~\cite{girimaji1990,Guala2005,Pumir2011}.  Since only the vorticity perpendicular to the rod contributes to its rotation rate, aligned particles have greatly reduced rotation rates.

Figure~\ref{Fig:varaincep2}b shows the dependence of rotation rate on aspect ratio for several simple cases in order to better understand the ways the data in Figure~\ref{Fig:varaincep2}a depends on both alignment with the velocity gradient tensor and the time evolution of the velocity gradient tensor.
If the velocity gradients are sampled from DNS but are delta-correlated in time, the particles do not develop alignment with the velocity gradients, and therefore the rotation rate variance matches the prediction for randomly oriented rods in  Eq.~\ref{Eq:p2var}.  When the velocity gradients are sampled from the DNS and held fixed in time, the alignment effect is too strong and the rotation rates are smaller than the rotation rates of rods advected in the turbulence at almost all aspect ratios.   Finally, we integrated particles through time independent plane shear flow which generates Jeffery orbits.   None of these cases has even qualitative agreement with the rotation rates of rods advected in the turbulence.  Since the dependence of the rotation rate variance on aspect ratio is sensitive to both instantaneous statistics and temporal correlations of the velocity gradient tensor, it provides a metric for evaluating models of the velocity gradients in turbulence that is directly accessible experimentally.
Figure~\ref{Fig:varaincep2}c shows the dependence of the rotation rate on the Reynolds number.  In simulations ranging from $R_\lambda=31$ to $R_\lambda=400$, the rotation rate variance changes by less than 9\% with the largest differences occurring for disks ($\alpha << 1$).
\begin{figure}[tb]
 \centerline{\includegraphics[scale=1]{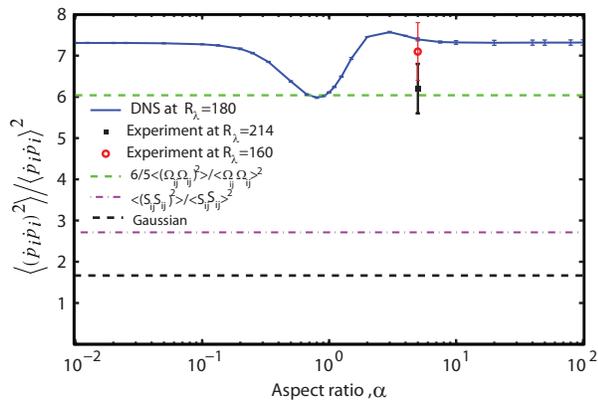}}
\caption{Fourth moment of the rotation rate of ellipsoids. Blue $+$ is the DNS at $R_\lambda =180$. Black $\blacksquare$ is the experiment at $R_\lambda =214$.   Red $\circ$ is the experiment at $R_\lambda =160 $. Green dashed line is the prediction for spheres uncorrelated with the direction of $\Omega_{ij}$. The purple dashed-dot line is the fourth moment of $S_{ij}$ and the black dashed line represents the case of a Gaussian distribution.
}
  \label{Fig:flatness}
  \end{figure}
For a more quantitative evaluation of the shape of the pdf of rotation rate squared, we report in Fig.~\ref{Fig:flatness} a normalized fourth moment of the rotation rate, $\langle (\dot{p}_i\dot{p}_i)^2 \rangle / \langle \dot{p}_i\dot{p}_i \rangle^2$ as a function of aspect ratio.   The experimental measurements are in fairly good agreement with the simulations considering that the differences are on the order of the experimental measurement errors and particle size may affect the tails of the experimental distribution as discussed above. The fourth moment for a sphere can be related to the fourth moment of the vorticity tensor by assuming that the orientation director is uncorrelated with $\Omega_{ij}$ which leads to
$ \langle (\dot{p}_i\dot{p}_i)^2 \rangle / \langle \dot{p}_i\dot{p}_i \rangle^2  = 6/5  \langle (\Omega_{ij}\Omega_{ij})^2 \rangle / \langle \Omega_{ij}\Omega_{ij} \rangle^2$, in good agreement with the simulations.  All these fourth moments are much larger than the value of 5/3 that is obtained if the components, $\dot{p}_i$, have a gaussian distribution. Rods ($\alpha >>1$) and disks ($\alpha << 1$) have nearly identical normalized fourth moments, and the variation with aspect ratio is less than 20\%, indicating that the dependence on particle shape of the normalized probability distribution is much weaker than that of the variance of the rotation rate in Fig~\ref{Fig:varaincep2}.

From both experiments and numerical simulations we have obtained a phenomenology of the rotational dynamics of anisotropic particles in turbulent fluid flow.  Rod-like particles have a rotation rate that is strongly affected by the alignment of the rods with the vorticity vector.  Disk-like particle also show effects of alignment although their rotation rate variance is closer to the randomly oriented case.  The transition between the two limiting aspect ratios is quite abrupt, with 80\% of the difference in rotation rate variance between disks and rods occurring between aspect ratios 0.5 and 2.  Thus for many bacteria or ice crystals that are transported in turbulent flows, the picture of a thin Lagrangian rod or disk is much better than the approximation that they are spheres.

We acknowledge support from COST Action MP0806: \textit{Particles in turbulence} and NSF grant DMR-0547712.  We thank Nicholas Ouellette for contributions to the early stages of this work.

%
%

\end{document}